\documentclass{moriond}

\def\be{\begin{equation}}
\def\ee{\end{equation}}
\def\bea{\begin{eqnarray}}
\def\eea{\end{eqnarray}}

\newcommand{\Photo}{}

\begin{document}
\vspace*{4cm}
\title{CHARGED CURRENT ANOMALIES WITH BARYONS}

\author{ D.BECIREVIC, F. JAFFREDO }

\address{
IJCLab, Pôle Théorie (Bât. 210), CNRS/IN2P3 et Université Paris-Saclay, 91405 Orsay, France}

\maketitle\abstracts{Motivated by the first observation of $\Lambda_b\to\Lambda_c\:\tau\bar{\nu}$ by LHCb, we discuss various observables that can be derived from angular distribution of this decay which, if measured, could be essential for explanation of the $b\to c\tau\bar{\nu}$ anomalies. Furthermore, we show that the measurement of $R_{\Lambda_c}=\mathcal{B}(\Lambda_b\to\Lambda_c\tau\bar{\nu})/\mathcal{B}(\Lambda_b\to\Lambda_c\mu\bar{\nu})$ does not change the current global picture arising from $R_D^{\rm}$ and $R_{D^*}^{\rm}$ being larger than predicted in the Standard Model. Finally, we show the example of a few angular observables, with and without a secondary decay, that could have a strong discriminating power on New Physics.}

\section{Introduction}

Hints of lepton flavor universality violation (LFUV) have been observed in $B$ meson decay by measuring the universality ratios $R_{D^{(*)}}$ and $R_{K^{(*)}}$\cite{1852846,Bordone:2016gaq,Abdesselam:2019dgh,Amhis:2019ckw}. Being the simplest in terms of valence quark content, mesons are hadrons of choice for the precision flavor physics observables both in experimental studies, due to their abundance in the clean environment of $B$-factories, and because of the relatively small theoretical uncertainties associated with the relevant hadronic matrix elements.

In what follows, we discuss the possibility of observing other hints of LFUV by looking at the baryon decays. Those decays involve the same $b\to c\ell\bar{\nu}$ semileptonic transition and are produced in large quantity at LHCb\cite{LHCb:2022piu}.


\section{Theory prediction}
\label{sec:theory}

To describe the charged current mediated semileptonic decay, in addition to the Standard Model (SM) Lagrangian, we also consider the following New Physics contributions (NP):
\begin{eqnarray}
\mathcal{L}_{\rm NP} &\supset& \sqrt{2}G_FV_{ij}\bigg[g_V(\overline{u}_i\gamma_\mu d_j)(\overline{\ell}_L\gamma^\mu\nu_L)+g_A(\overline{u}_i\gamma_\mu\gamma_5d_j)(\overline{\ell}_L\gamma^\mu\nu_L)\\\nonumber
&&+g_S(\overline{u}_id_j)(\overline{\ell}_R\nu_L)+g_P(\overline{u}_i\gamma_5d_j)(\overline{\ell}_R\nu_L)+g_T(\overline{u}_i\sigma_{\mu\nu}(1-\gamma_5)d_j)(\overline{\ell}_R\sigma^{\mu\nu}\nu_L)\bigg]+{\rm h.c.}
\end{eqnarray}

The resulting $\mathcal{L}=\mathcal{L}_{\rm SM}+\mathcal{L}_{\rm NP}$ can describe various flavor transitions; including the semileptonic decays of baryons such as $\Lambda_b\to\Lambda_c\:\ell\nu$, $\Lambda_b\to\Lambda_c^*\ell\nu$, $\Lambda_c\:\to\Lambda\ell\nu$, $\Lambda\to p\ell\nu$. The main obstacles to the precision computation of decay rates are the hadronic matrix element of the quark currents sandwiched by the two hadronic states. Such matrix elements have been computed to a remarkable accuracy for many processes by means of lattice QCD \cite{Detmold:2015aaa,Meinel:2021mdj,Datta:2017aue}. For baryons, the results are expressed in term of 10 form factors, denoted $F_0, F_+, F_\perp, G_0, G_+, G_\perp, h_+,h_\perp, \widetilde{h}_+,\widetilde{h}_\perp$ (resp. $3$ vector, $3$ axial, and $4$ tensor form factors). The tensor form factors are not needed for the SM prediction but they are required in some scenarios of NP. The fact that all of these $10$ form factors are available from lattice QCD for the $\Lambda_b\to\Lambda_c\:\ell\nu$ decay in the full phase space is a unique situation that makes this study particularly interesting.

By focusing on $\Lambda_b\to\Lambda_c\:\ell\nu$, we can express the decay distribution as:
\begin{equation}
\frac{{\rm d}^2\Gamma(\Lambda_b\to \Lambda^{\lambda_c}_c\ell^{\lambda_l}\nu)}{{\rm d}q^2{\rm d}\cos\theta}={ a^{\lambda_l}_{\lambda_c}}(q^2)+{b_{\lambda_c}^{\lambda_l}}(q^2){\cos\theta}+{c_{\lambda_c}^{\lambda_l}}(q^2){\cos^2\theta},
\label{eq:angularDistr}
\end{equation}
where $\lambda_c$ and $\lambda_l$ are the polarization states of the outgoing $\Lambda_c$ and $\ell$. $\theta$ is the angle between the lepton and the $\Lambda_c$ baryon in the dilepton rest-frame. The angular coefficients $a$, $b$, and $c$ are functions of $q^2=(p_{\Lambda_b}-p_{\Lambda_c})^2=(p_\ell+p\nu)^2$, the NP couplings $g_{i},\:(i\in V,A,S,P,T)$, form factors and the polarizations of the outgoing particles. Explicit formulas can be found in Refs.\cite{Datta:2017aue,Boer:2019zmp,Habyl:2015xka,Mu:2019bin,Penalva:2020ftd}, as well as in Ref.\cite{Becirevic:2022bev}, in which we discuss this decay in more detail.

From the explicit formulas, we can see that
\begin{equation}
b^-_\pm= \pm (a^-_\pm+c^-_\pm).
\label{eq:linearConstraint}
\end{equation}
Contrary to the case of the pseudoscalar to pseudoscalar meson decays where we only have a single spin-$1/2$ particle in the final state yielding at most $4$ linearly independent observables\cite{Angelescu:2021lln}, here we expect $12$ observables ($4$ for each of the $a,b,c$). Eq.~(\ref{eq:linearConstraint}) reduces this number to $10$.

Setting all $g_i=0$, we can compute the SM prediction for $\Lambda_b\to\Lambda_c\:\tau\nu$ and $\Lambda_b\to\Lambda_c\:\mu\nu$, and combine them to
\begin{equation}
{R_{\Lambda_c}}=\frac{\mathcal{B}(\Lambda_b\to\Lambda_c\:\tau\nu)}{\mathcal{B}(\Lambda_b\to\Lambda_c\:\mu\nu)},
\end{equation}
in which a bulk of uncertainties on the hadronic matrix elements cancel. Using the lattice QCD form factors\cite{Detmold:2015aaa,Datta:2017aue}, we get
\begin{equation}
R_{\Lambda_c}^{{\rm SM}}=0.333(13).
\end{equation}

\section{New LHCb measurement}
\label{sec:experiment}

LHCb recently reported on the first observation of the semileptonic decay $\Lambda_b\to\Lambda_c\:\tau\nu$.\cite{LHCb:2022piu} They combine it with the old LEP result for $\mathcal{B}(\Lambda_b\to\Lambda_c\:\mu\nu)$\cite{DELPHI:2003qft}, and obtain 	
\begin{equation}
R_{\Lambda_c}^{{\rm LHCb}}=0.242(76).
\end{equation}
The error on this measurement is dominated by the systematics coming from the external branching fractions: (i) $\mathcal{B}(\Lambda_b\to\Lambda_c\:\mu\nu)$\cite{DELPHI:2003qft} ($15\%$), and (ii) from the normalization channel $\mathcal{B}(\Lambda_b\to\Lambda_c \:3\pi)$\cite{CDF:2011aa,LHCb:2011poy} ($23\%$).

This result is compatible with the SM prediction, merely $\sim 1\sigma$ below. Therefore the anomaly observed with mesons ($R_{D^{(*)}}^{\rm Exp}>R_{D^{(*)}}^{\rm SM}$) is not yet observed with baryons, which might change once the more accurate LHCb data become available. We remind the reader that the observed value of $R_D$ and $R_{D^*}$ are respectively $1.5\sigma$ and $3.5\sigma$ above their SM predictions \cite{Abdesselam:2019dgh,Amhis:2019ckw,Becirevic:2020rzi}. It can be shown that due to the relatively large uncertainty, $R_{\Lambda_c}$ has a small effect on the overall constraints on NP.

To illustrate the impact of this new measurement, we consider several scenarios in which NP only involves a single NP coupling (or a combination of two couplings). In particular, we consider scenarios involving $g_{V_L}=(g_V-g_A)/2$ and $g_{S_L}=(g_S-g_P)/2$, since they can be observed in several popular extensions of the SM. We also show the results for $g_T$ only and for the combination $g_{S_L}=\pm 4g_T$, which is usually evoked in the scenarios involving Low Energy Scalar Leptoquarks \cite{Becirevic:2020rzi}. Since $g_{S_L}$ and $g_T$ are renormalization scale dependant, after running from the matching scale $\mu\simeq 1\:{\rm TeV}$ down to $\mu=m_b$, the above relations become $g_{S_L}=-8.5\:g_T$ and $g_{S_L}=8.1\:g_T$ \cite{Gonzalez-Alonso:2017iyc}. In the latter scenario we consider only the imaginary values of $g_{S_L}$ since it has been shown that a real-valued $g_{S_L}$ is not compatible with the observed values of $R_D$ and $R_{D^*}$\cite{Becirevic:2020rzi}.

We extracted the preferred value ($1\sigma$ limit) of the NP couplings for the aforementioned scenarios, collected in Tab.~\ref{tab:WCfits}. As expected, the values obtained with $R_{\Lambda_c}$ only are compatible with zero, but the inclusion of $R_{\Lambda_c}$ in the fit does not sensibly change the values of $g_i$ with respect to the values obtained using $R_D$ and $R_{D^*}$.

\begin{table}[h]
\centering
{\renewcommand{\arraystretch}{1.8}
\begin{tabular}{|c|cccc|}
\hline
Wilson Coefficient              & $R_D$ and $R_{D^*}$ & $R_{\Lambda_c}$  & Combined            & $\chi^2_{\rm min}/{\rm d.o.f}$ \\ \hline\hline
$g_{V_L}$                       & $0.084\pm 0.029$    & $-0.15\pm 0.14$ & $0.077\pm 0.035$    & $0.06\to 1.3$ \\
$g_{S_L}$                       & $-1.47\pm 0.08$     & $-0.53\pm 0.54$ & $-1.45\pm 0.11$     & $0.5\to 2.1$ \\
$g_T$                           & $-0.027\pm 0.011$   & $0.13\pm 0.14$  & $-0.026\pm 0.013$   & $1.2\to 1.7$ \\
$g_{S_L}=+4g_T\in i {\rm I\!R}$ & $\pm 0.49 \pm 0.10$ & $0.0\pm 0.39$   & $\pm 0.47\pm 0.13$  & $0.9\to 1.6$ \\
$g_{S_L}=-4g_T$                 & $0.16 \pm 0.06$     & $0.0\pm 0.39$   & $0.15\pm 0.07$      & $0.7\to 1.0$ \\ \hline
\end{tabular}}
\caption{Fits to single Wilson coefficients based on $R_D$ and $R_{D^*}$, $R_{\Lambda_c}$ alone, or all $3$ observables.}
\label{tab:WCfits}
\end{table}

\begin{figure}[h]
\begin{minipage}{0.49\linewidth}
\centerline{\includegraphics[width=0.98\linewidth]{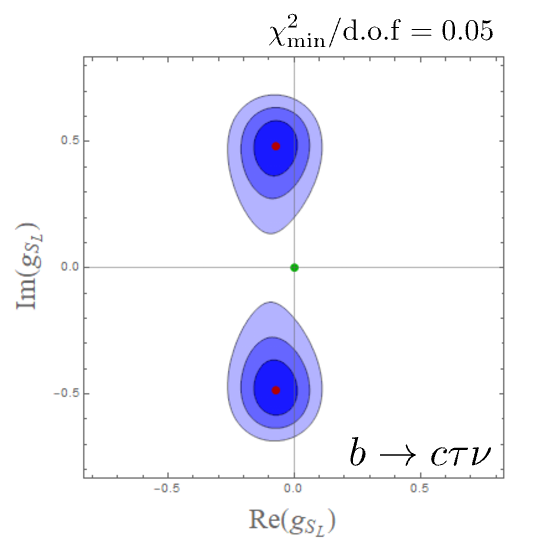}}
\end{minipage}
\hfill
\begin{minipage}{0.49\linewidth}
\centerline{\includegraphics[width=0.98\linewidth]{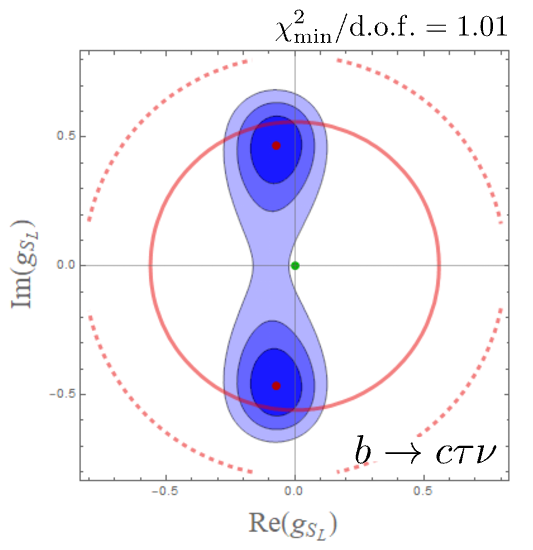}}
\end{minipage}
\caption{Fit to the complex Wilson coefficient $g_{S_L}$ in the scenario $g_{S_L}=4g_T$ using only the constraints from $R_D$ and $R_{D^*}$ (left) and with adding the constraints from $R_{\Lambda_c}$. The constraint from $R_{\Lambda_c}$ alone is also shown at $1$ and $2\sigma$ in red. The SM is represented by the green point.}
\label{fig:avantapres}
\end{figure}

The effect of inclusion of $R_{\Lambda_c}^{\rm exp}$ in the scenario with complex $g_{S_L}=4g_T$ is illustrated in Fig.~\ref{fig:avantapres}. We see that the solutions for $g_{S_L}$ to $2\sigma$, as derived from $R_D^{\rm exp}$ and $R_{D^*}^{\rm exp}$ (left panel) remain unchanged (right panel). Only the $\chi^2_{\rm min}$ value gets a slight shift. Instead, the $3\sigma$ region, after including $R_{\Lambda}^{\rm exp}$, allows for a pure real solution to $g_{S_L}$, which nevertheless is not compatible with SM.

\section{Angular observables and secondary decay}
\label{sec:angular}

LHCb measured one of the $10$ observables mentioned in Sect.~\ref{sec:theory}, nameley $\mathcal{B}(\Lambda_b\to\Lambda_c\tau\bar{\nu})$. By a simple inspection of Eq.~(\ref{eq:angularDistr}), one can immediately see that any physics information contained in $b(q^2)$ is lost when integrating over $\cos\theta$. To regain access to this information one considers angular observables. In particular, the Forward-Backward Asymmetry ($\mathcal{A}_{\rm fb}$) defined as (we neglect spin indices when summing over all polarization)
\begin{eqnarray}
\mathcal{A}_{\rm fb}(q^2)&=&\frac{1}{\Gamma_{\rm tot}}\bigg(\int_0^1 - \int_{-1}^0\bigg)  \frac{{\rm d}\Gamma(\Lambda_b\to \Lambda_c\ell\nu)}{{\rm d}q^2{\rm d}\!\cos\theta}{\rm d}\!\cos\theta=\frac{b(q^2)}{\Gamma_{\rm tot}},\\ 
\left\langle\mathcal{A}_{\rm fb}\right\rangle &=& \int \mathcal{A}_{\rm fb}(q^2){\rm d}q^2.
\end{eqnarray}
Similarly, we can individually access $c(q^2)$ by looking at the so-called ``convexity" of the distribution\cite{Angelescu:2021lln}:
\begin{eqnarray}
\mathcal{A}_{\pi/3}(q^2)&=&\frac{1}{2\Gamma_{\rm tot}}\bigg(\int_{1/2}^1 + \int_{-1}^{-1/2} - \int_{-1/2}^{1/2}\bigg) \frac{{\rm d}\Gamma(\Lambda_b\to \Lambda_c\ell\nu)}{{\rm d}q^2{\rm d}\!\cos\theta}{\rm d}\!\cos\theta = \frac{c(q^2)}{2\Gamma_{\rm tot}},\\
\left\langle\mathcal{A}_{\pi/3}\right\rangle &=& \int \mathcal{A}_{\pi/3}(q^2){\rm d}q^2.
\end{eqnarray}
Notice that $A_{\pi/3}$ is obtained experimentally by counting positively events where the outgoing lepton falls in the cones of radius $\pi/3$ with respect to the direction of the outgoing baryon, and negatively otherwise.\\

One possibility to access many more angular observables is to consider the secondary baryon decay. Restricting ourselves to only 2-body decays, there are $2$ possibilities $\Lambda_c\to\Lambda\pi$ and $\Lambda_c\to pK_S$ (the branching fraction of which are $1.30(7)\%$ and $1.59(8)\%$ respectively). The small branching fraction of the secondary decay certainly reduces the statistics, since the current measurement is made using the reconstruction channel $\Lambda_c\to p K^-\pi^+$ for which $\mathcal{B}=6.28(32)\%$ \cite{LHCb:2022piu}.

One can express the total angular distribution by introducing only $2$ new parameters: the branching fraction of the secondary decay and an asymmetry parameter $\alpha$,
\begin{equation}
\alpha=\frac{\left\langle \Lambda^+\pi|\Lambda_c^+\right\rangle^2-\left\langle\Lambda^-\pi|\Lambda_c^- \right\rangle^2}{\left\langle \Lambda^+\pi|\Lambda_c^+ \right\rangle^2+\left\langle \Lambda^-\pi|\Lambda_c^-\right\rangle^2},
\end{equation}
and similarly for $\Lambda_c\to pK_S$. The parameter $\alpha$ has only been precisely measured for $\Lambda_c\to\Lambda\pi$, which is why we opt for that secondary decay, even though not every angular observable depends on $\alpha$.

\begin{figure}[h]
\begin{minipage}{0.49\linewidth}
\centerline{\includegraphics[width=0.98\linewidth]{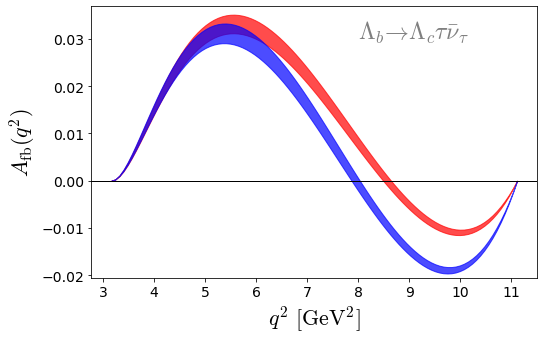}}
\end{minipage}
\hfill
\begin{minipage}{0.49\linewidth}
\centerline{\includegraphics[width=0.98\linewidth]{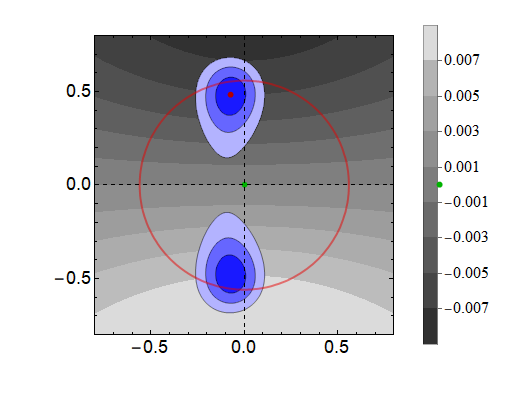}}
\end{minipage}
\caption{(a): Forward-Backward Asymmetry before integration over $q^2$ in the SM (blue) and for the best fit point in the scenario $g_{S_L}=4g_T$ (red). In the SM, we observe a cancellation between the low and high $q^2$, leading to a good sensitivity of this observable.$\quad$(b): Fit to the Wilson coefficient $g_{S_L}$ in the scenario $g_{S_L}=4g_T$ as obtain in Fig.~\ref{fig:avantapres}. In the background is plotted the value of $D_4$, which exhibits a vertical gradient due to its dependence to the imaginary part of $g_{S_L}$.}
\label{fig:angularObs}
\end{figure}

The full angular distribution for $\Lambda_b\to\Lambda_c(\to \Lambda\pi)\:\tau\nu$ involves two new angles: $\theta_\Lambda$ (between $\Lambda_c$ and $\Lambda$ in the $\Lambda\pi$ rest frame) and $\phi$ (between $\Lambda\pi$- and $\ell\nu$-planes). The expression similar to Eq.~(\ref{eq:angularDistr}) now writes:
{\setlength\arraycolsep{1.4pt}
\begin{eqnarray}\nonumber
\frac{d^4\Gamma^{\lambda_l}}{dq^2d\cos\theta d \cos\theta_\Lambda d\phi} &=& A_1^{\lambda_l}+A_2^{\lambda_l}\cos\theta_\Lambda +\left(B_1^{\lambda_l}+B_2^{\lambda_l}\cos\theta_\Lambda\right)\cos\theta+\left(C_1^{\lambda_l}+C_2^{\lambda_l}\cos\theta_\Lambda\right)\cos^2\theta\\
&+&\left(D_3^{\lambda_l}\sin\theta_\Lambda\cos\phi+D_4^{\lambda_l}\sin\theta_\Lambda\sin\phi\right)\sin\theta\\\nonumber
&+&\left(E_3^{\lambda_l}\sin\theta_\Lambda\cos\phi+E_4^{\lambda_l}\sin\theta_\Lambda\sin\phi\right)\sin\theta\cos\theta.
\end{eqnarray}}
The expressions for the $q^2$-dependent angular coefficients $A_{1,2},B_{1,2},C_{1,2},D_{3,4}$ and $E_{3,4}$ can be found in in Refs.\cite{Datta:2017aue,Boer:2019zmp,Habyl:2015xka,Mu:2019bin,Penalva:2020ftd}. The important points can be summarized as follows:
\begin{itemize}
\item We can have a Forward-Backward Asymmetry with respect to the lepton (yielding $B_1$), the baryon (yielding $A_2$ and $C_2$), or both at the same time (yielding $B_2$).
\item The lepton Forward-Backward Asymmetry $\left\langle\mathcal{A}_{\rm fb}\right\rangle$ has a small positive value in the SM due to the cancellation between the low and high $q^2$ regions. It can thus be very sensitive to NP, as illustrated in Fig.~\ref{fig:angularObs}a, because the point at which $\mathcal{A}_{fb}(q^2)=0$ is shifted with respect to the SM value.
\item There is no equivalent of the convexity $\mathcal{A}_{\pi/3}$ for the baryon since there is no $\cos^2\theta_\Lambda$ term. 
\item The spin of the outgoing $\Lambda$ baryon does not bring any new information beyond $\alpha$. In other words, the secondary decay encodes information on the polarization into angular observables and it can actually be used to extract $\alpha$.  
\item In addition to $A$, $B$, and $C$, we have other observables sensitive to $\phi$. In particular, $E_4$ and $D_4$ are proportional to the imaginary part of the Wilson coefficients, and thus they are exactly zero in the SM and the scenarios of NP without a new CP-violating phase. Therefore their measurement could be a strong zero-test of the NP phase. An example of a scenario with non-zero $D_4$ is shown in Fig.~\ref{fig:angularObs}b.
\end{itemize}
The angular observables exhibit a desired features of cancellation of the CKM elements, a partial cancellation of hadronic uncertainties, and not requiring any external branching fraction. They are sensitive to various combinations of NP couplings and form factors and thus could help discriminate among various scenarios of NP. It is also very important to note that one can observe discrepancies in angular observables with respect to their SM predictions even if $R_{\Lambda_c}$ is perfectly SM-like.

\section{Conclusion}

In addition to the semileptonic decays of $B$ mesons, $\Lambda_b$ baryons offer interesting opportunities to look for the presence of NP, beyond the SM. In particular, the measurement of the universality ratio $R_{\Lambda_c}$ can provide an independent test of LFUV since the relevant hadronic uncertainties are well under control, thanks to Lattice QCD.

The first measurement of $R_{\Lambda_c}$ by LHCb was an important feasibility test. The reported value is compatible with the SM prediction but also with the previously preferred scenarios of NP selected by $R_{D^{(*)}}^{\rm Exp}$. We updated those scenarios including the new observable. $R_{\Lambda_c}^{\rm Exp}$ is currently weaker a constraint than those obtained by $R_D^{\rm Exp}$ and $R_{D^*}^{\rm Exp}$, but could quickly become competitive since the sensitivity is held back by the systematics in the normalization channel.

We also explored various angular observables to NP, both for the $\Lambda_b\to\Lambda_c\:\tau\nu$ decay and when including the secondary decay $\Lambda_c\to\Lambda\pi$, and found interesting properties for many of them. If measured, these observables could help disentangling among various scenarios of physics beyond the SM.


\section*{References}

\begin{thebibliography}{99}


\bibitem{1852846}
R.~Aaij \textit{et al.} [LHCb],
[arXiv:2103.11769 [hep-ex]].

\bibitem{Bordone:2016gaq}
M.~Bordone, G.~Isidori and A.~Pattori,
Eur. Phys. J. C \textbf{76} (2016) no.8, 440
[arXiv:1605.07633 [hep-ph]];
G.~Isidori, S.~Nabeebaccus and R.~Zwicky,
JHEP \textbf{12} (2020), 104
[arXiv:2009.00929 [hep-ph]].

\bibitem{Abdesselam:2019dgh}
A.~Abdesselam \textit{et al.} [Belle],
[arXiv:1904.08794 [hep-ex]].

\bibitem{Amhis:2019ckw}
Y.~S.~Amhis \textit{et al.} [HFLAV],
[arXiv:1909.12524 [hep-ex]];

\bibitem{LHCb:2022piu}
R.~Aaij et~al.
\newblock {\em Phys. Rev. Lett.}, 128(19):191803, 2022.

\bibitem{Detmold:2015aaa}
W.~Detmold, C.~Lehner, and S.~Meinel.
\newblock {\em Phys. Rev. D}, 92(3):034503, 2015.

\bibitem{Meinel:2021mdj}
S.~Meinel and G.~Rendon.
\newblock {\em Phys. Rev. D}, 105(5):054511, 2022.

\bibitem{Datta:2017aue}
A.~Datta, S.~Kamali, S.~Meinel, and A.~Rashed.
\newblock {\em JHEP}, 08:131, 2017.

\bibitem{Boer:2019zmp}
P.~B\"oer, A.~Kokulu, J.-N. Toelstede, and D.~van Dyk.
\newblock {\em JHEP}, 12:082, 2019.

\bibitem{Habyl:2015xka}
N.~Habyl, T.~Gutsche, M.~A. Ivanov, J.~G. K\"orner, V.~E. Lyubovitskij, and
  P.~Santorelli.
\newblock {\em Int. J. Mod. Phys. Conf. Ser.}, 39:1560112, 2015.

\bibitem{Mu:2019bin}
X.-L. Mu, Y.~Li, Z.-T. Zou, and B.~Zhu.
\newblock {\em Phys. Rev. D}, 100(11):113004, 2019.

\bibitem{Penalva:2020ftd}
N.~Penalva, E.~Hern\'andez, and J.~Nieves.
\newblock {\em Phys. Rev. D}, 102(9):096016, 2020.

\bibitem{Becirevic:2022bev}
D.~Be\v{c}irevi\'c and F.~Jaffredo,
[arXiv:2209.13409 [hep-ph]].

\bibitem{Angelescu:2021lln}
A.~Angelescu, D.~Be\v{c}irevi\'c, D.~A. Faroughy, F.~Jaffredo, and
  O.~Sumensari.
\newblock {\em Phys. Rev. D}, 104(5):055017, 2021.

\bibitem{DELPHI:2003qft}
J.~Abdallah et~al.
\newblock {\em Phys. Lett. B}, 585:63--84, 2004.

\bibitem{CDF:2011aa}
T.~Aaltonen et~al.
\newblock {\em Phys. Rev. D}, 85:032003, 2012.

\bibitem{LHCb:2011poy}
R.~Aaij et~al.
\newblock {\em Phys. Rev. D}, 84:092001, 2011.
\newblock [Erratum: Phys.Rev.D 85, 039904 (2012)].

\bibitem{Becirevic:2020rzi}
D.~Be\v{c}irevi\'c, F.~Jaffredo, A.~Pe\~nuelas, and O.~Sumensari.
\newblock {\em JHEP}, 05:175, 2021.

\bibitem{Gonzalez-Alonso:2017iyc}
M.~Gonz\'alez-Alonso, J.~Martin~Camalich, and K.~Mimouni.
\newblock {\em Phys. Lett. B}, 772:777--785, 2017.


\end{thebibliography}
\end{document}